\documentclass[sigconf,pbalance,nonacm,authorversion]{acmart}
\copyrightyear{2026}
\acmYear{2026}
\setcopyright{cc}
\setcctype{by}
\acmConference[ITiCSE 2026]{Proceedings of the 31st ACM Conference on Innovation and Technology in Computer Science Education V. 1}{July 10--15, 2026}{Madrid, Spain}
\acmBooktitle{Proceedings of the 31st ACM Conference on Innovation and Technology in Computer Science Education V. 1 (ITiCSE 2026), July 10--15, 2026, Madrid, Spain}
\acmDOI{10.1145/3803400.3809322}
\acmISBN{979-8-4007-2634-7/2026/07}

\usepackage{enumitem}
\usepackage[most]{tcolorbox}
\usepackage{csquotes}
\usepackage{xspace}

\usepackage[nolist]{acronym}
\begin{acronym}

    \acro{3DES}{Triple-DES}
    \acro{4CID}[4C/ID]{Four-Component Instructional Design}
    \acro{12P}[12P]{12 Principles of Computing Pedagogy}

    \acro{ACT-R}{Adaptive Control of Thought-Rational}
    \acro{AES}{Advanced Encryption Standard}
    \acro{ALU}{Arithmetic Logic Unit}
    \acro{ANOVA}{Analysis of Variance}
    \acroplural{ANOVA}[ANOVAs]{Analyses of Variance}
    \acro{API}{Application Programming Interface}
    \acro{AOI}{Area of Interest}
    \acroplural{AOI}[AOIs]{Areas of Interest}
    \acro{ARX}{Addition Rotation XOR}
    \acro{ATPG}{Automatic Test Pattern Generation}
    \acro{ASIC}{Application Specific Integrated Circuit}
    \acro{ASIP}{Application Specific Instruction-Set Processor}
    \acro{AS}{Active Serial}
    
    \acro{BDD}{Binary Decision Diagram}
    \acro{BGL}{Boost Graph Library}
    \acro{BKA}{German Federal Criminal Police Office}
    \acro{BNF}{Backus-Naur Form}
    \acro{BRAM}{Block-Ram}
    
    \acro{CBC}{Cipher Block Chaining}
    \acro{CFB}{Cipher Feedback Mode}
    \acro{CFG}{Control Flow Graph}
    \acro{CLB}{Configurable Logic Block}
    \acro{CLI}{Command Line Interface}
    \acro{CMOS}{Complementary Metal-Oxide-Semiconductor}
    \acro{COFF}{Common Object File Format}
    \acro{CPA}{Correlation Power Analysis}
    \acro{CPU}{Central Processing Unit}
    \acro{CRediT}{Contributor Roles Taxonomy}
    \acro{CRC}{Cyclic Redundancy Check}
    \acro{CTA}{Concurrent Think Aloud}
    \acro{CTR}{Counter}
    
    \acro{DAA}{Disassemble / Analyze / Assemble}
    \acro{DC}{Direct Current}
    \acro{DES}{Data Encryption Standard}
    \acro{DFA}{Differential Frequency Analysis}
    \acro{DFT}{Discrete Fourier Transform}
    \acro{DIP}{Distinguishing Input Pattern}
    \acro{DLL}{Dynamic Link Library}
    \acro{DMA}{Direct Memory Access}
    \acro{DNF}{Disjunctive Normal Form}
    \acro{DPA}{Differential Power Analysis}
    \acro{DSO}{Digital Storage Oscilloscope}
    \acro{DSP}{Digital Signal Processing}
    \acro{DUT}{Design Under Test}
    
    \acro{ECB}{Electronic Code Book}
    \acro{ECC}{Elliptic Curve Cryptography}
    \acro{EEPROM}{Electrically Erasable Programmable Read-only Memory}
    \acro{EDA}{Electronic Design Automation}
    \acro{EMA}{Electromagnetic Emanation}
    \acro{EM}{electro-magnetic}
    \acro{EMME}{Eye Movement Modeling Example}
    \acro{EU}{European Union}
    
    \acro{FFT}{Fast Fourier Transformation}
    \acro{FF}{Flip Flop}
    \acro{FI}{Fault Injection}
    \acro{FIR}{Finite Impulse Response}
    \acro{FPGA}{Field Programmable Gate Array}
    \acro{FSM}{Finite State Machine}
    
    \acro{GDS}{Graphic Design System}
    \acro{GLMM}{Generalized Linear Mixed Model}
    \acro{GLM}{Generalized Linear Model}
    \acro{GT}{Grounded Theory}
    \acro{GUI}{Graphical User Interface}
    
    \acro{HCI}{Human Computer Interaction}
    \acro{HDL}{Hardware Description Language}
    \acro{HD}{Hamming Distance}
    \acro{HF}{High Frequency}
	\acro{HRE}{Hardware Reverse Engineering}
    \acro{HSM}{Hardware Security Module}
    \acro{HW}{Hamming Weight}
    
    \acro{IC}{Integrated Circuit}
    \acro{ICC}{Intraclass Correlation}
    \acro{IO}[I/O]{Input/Output}
    \acro{IOB}{Input Output Block}
    \acro{IoT}{Internet of Things}
    \acro{IRB}{Institutional Review Board}
    \acro{IP}{Intellectual Property}
    \acro{IPIP}{International Personality Item Pool}
    \acro{IPIP-NEO}{International Personality Item Pool Representation of the NEO PI-R}
    \acro{IQ}{Intelligence Quotient}
    \acro{IRR}{Inter-Rater Reliability}
    \acro{ISA}{Instruction Set Architecture}
    \acro{ISCED}{International Standard Classification of Education}
    \acro{ITS}{Intelligent Tutoring System}
    \acro{IV}{Initialization Vector}
    
    \acro{JTAG}{Joint Test Action Group}
    
    \acro{KAT}{Known Answer Test}
    \acro{KLI}{Knowledge-Learning-Instruction}

    \acro{LFSR}{Linear Feedback Shift Register}
    \acro{LMM}{Linear Mixed Model}
    \acro{LSB}{Least Significant Bit}
    \acro{LUT}{Look-up table}
    
    \acro{MAC}{Message Authentication Code}
    \acro{MAD}{Median Absolute Deviation}
    \acro{MIPS}{Microprocessor without Interlocked Pipeline Stages}
    \acro{ML}{Machine Learning}
    \acro{MLM}{Multilevel Model}
    \acro{MLR}{Multiple Linear Regression}
    \acro{RMMR}{Repeated Measures Multiple Regression}
    \acro{MMIO}{Memory Mapped \acl{IO}}
    \acro{MSB}{Most Significant Bit}
    
    \acro{NASA}{National Aeronautics and Space Administration}
    \acro{NRE}{Netlist Reverse Engineering}
    \acro{NCT}{Number Connection Task}
    \acro{NSA}{National Security Agency}
    \acro{NVM}{Non-Volatile Memory}
    
    \acro{OEM}{Original Equipment Manufacturer}
    \acro{OFB}{Output Feedback Mode}
    \acro{OISC}{One Instruction Set Computer}
    \acro{ORAM}{Oblivious Random Access Memory}
    \acro{OS}{Operating System}
    \acro{OSF}{Open Science Framework}
    
    \acro{PAR}{Place-and-Route}
    \acro{PCB}{Printed Circuit Board}
    \acro{PC}{Personal Computer}
    \acro{PDK}{Process Design Kit}
    \acro{PS}{Processing Speed}
	\acro{PR}{Perceptual Reasoning}
    \acro{PUF}{Physically Unclonable Function}

    \acro{QA}{Quality Assurance}
    
    \acro{RISC}{Reduced Instruction Set Computer}
    \acro{RNG}{Random Number Generator}
    \acro{ROBDD}{Reduced Ordered Binary Decision Diagram}
    \acro{ROM}{Read-Only Memory}
    \acro{ROP}{Return-oriented Programming}
    \acro{RTA}{Retrospective Think Aloud}
    \acro{RTL}{Register Transfer Level}
    \acro{RUB}{Ruhr University Bochum}
    
    \acro{SAT}{Boolean satisfiability}
    \acro{SEM}{Scanning Electron Microscope}
    \acro{SEMOBS}{Self-Modifying Bitstreams}
    \acro{SCA}{Side-Channel Analysis}
    \acro{SHA}{Secure Hash Algorithm}
    \acro{SNR}{Signal-to-Noise Ratio}
    \acro{SoC}{System-on-Chip}
    \acroplural{SoC}[SoCs]{Systems-on-Chip}
    \acro{SPA}{Simple Power Analysis}
    \acro{SPI}{Serial Peripheral Interface Bus}
    \acro{SRAM}{Static Random Access Memory}
    \acro{SRE}{Software Reverse Engineering}
    \acro{STI}{shallow trench isolation}
    
    \acro{TA}{Think Aloud}
    \acro{TRNG}{True Random Number Generator}
    
    \acro{UART}{Universal Asynchronous Receiver Transmitter}
    \acro{UHF}{Ultra-High Frequency}
    \acro{UI}{User Interface}
    
    \acro{vFPGA}{virtual FPGA}
    \acro{VC}{Verbal Comprehension}
    \acro{VLSI}{Very-Large-Scale Integration}
    
    \acro{WAIS-IV}{Wechsler Adult Intelligence Scale}
    \acro{WM}{Working Memory}
    \acro{WISC}{Writeable Instruction Set Computer}
    
    \acro{XDL}{Xilinx Description Language}
    \acro{XTS}{XEX-based Tweaked-codebook with ciphertext Stealing}

    \acro{ZITiS}{German Central Office for Information Technology in the Security Sector}
    \acro{ZVT}{Zahlen-Verbindungs-Test}    
\end{acronym}

\newif\iftodo
\todofalse

\newtcolorbox{lessonbox}{
  enhanced,
  frame hidden,
  colback=black!3,
  borderline west={0.8pt}{0pt}{black},
  sharp corners,
  before skip=6pt,
  after skip=3pt,
  boxsep=.8pt,
  left=4pt,right=2pt,top=1pt,bottom=1pt,
  toprule=0.01pt,
  bottomrule=0.01pt,
}

\newenvironment{lesson}{%
  \begin{lessonbox}\noindent\textbf{Lesson learned: }\ignorespaces
}{%
  \end{lessonbox}
}

\newcommand{\eg}{e.\,g.}

\newcommand{\etal}{\textit{et al.}\@\xspace}

\makeatletter
\newcommand{\blind}[1]{%
    \if@ACM@anonymous %
        \texttt{Blinded Authors}
    \else%
        #1%
    \fi%
}
\makeatother%

\begin{document}

%\title[Lessons from Eight Years in a Niche Cybersecurity Domain]{Designing a Hardware Reverse Engineering Course:\\Lessons from Eight Years in a Niche Cybersecurity Domain}
\title[Lessons from Eight Years in a Rapidly Evolving Tech Domain]{Designing a Hardware Reverse Engineering Course:\texorpdfstring{\\}{ }Lessons from Eight Years in a Rapidly Evolving Tech Domain}

% Alternative Suggestions
% Teaching Hardware Reverse Engineering: Lessons from Eight Years of Course Development
% Developing a Niche Cybersecurity Course: Lessons from Eight Years of Teaching Hardware Reverse Engineering
% Developing a Niche Cybersecurity Course: Lessons from Eight Years of Hardware Reverse Engineering Education
% Developing a Niche Cybersecurity Course in Hardware Reverse Engineering: Lessons from Eight Years


\author{Zehra Karada\u{g}}
%\authornote{First author and corresponding author.}
\orcid{0009-0001-6863-3184}
\affiliation{%
  \institution{Ruhr University Bochum}
    %\city{Bochum}
  %\country{Germany} \\
  \institution{Max Planck Institute\\ for Security and Privacy}
  \city{Bochum}
  %\state{NRW}
  \country{Germany}
}

\author{René Walendy}
\orcid{0000-0002-5378-3833}
\affiliation{%
  %\institution{MPI-SP}
  \institution{Max Planck Institute\\ for Security and Privacy} 
  \city{Bochum}
  %\state{NRW}
  \country{Germany}}

\author{Carina Wiesen}
\orcid{0000-0002-4403-1656}
\affiliation{%
  \institution{Ruhr University Bochum}
  \city{Bochum}
  %\state{NRW}
  \country{Germany}
}

\author{Christof Paar}
\orcid{0000-0001-8681-2277}
\affiliation{%
  \institution{Max Planck Institute\\ for Security and Privacy}
  %\institution{MPI-SP}
  \city{Bochum}
  %\state{NRW}
  \country{Germany}}

\author{Nikol Rummel}
\orcid{0000-0002-3187-5534}
\affiliation{%
  \institution{Ruhr University Bochum}
  \institution{Center for Advanced Internet Studies}
  \city{Bochum}
  %\state{NRW}
  \country{Germany}
}

\author{Steffen Becker}
\orcid{0000-0001-7526-5597}
\affiliation{%
  \institution{Ruhr University Bochum}
      %\city{Bochum}
  %\country{Germany} \\
  \institution{Max Planck Institute\\ for Security and Privacy}
  \city{Bochum}
  %\state{NRW}
  \country{Germany}
}

%\renewcommand{\shortauthors}{Karada\u{g} et al.}
\renewcommand{\shortauthors}{Zehra Karadağ et al.}
%% No italics, no superscripts, not anonymous
%% Use footnote or author note to identify equal contribution and/or contact author info


\begin{abstract} % max 250 words
% motivation, 
\acp{IC} are omnipresent, yet their globalized manufacturing process remains vulnerable to supply chain threats.
% problem statement
\ac{HRE} is essential for detecting such threats and re-establishing trust; however domain experts remain scarce due to a lack of educational programs.
% approach
To contribute educational insights in this critical and rapidly evolving technology domain, we present our \ac{HRE} course focusing on digital circuit analysis and digital circuit extraction from \acp{IC}.
The course targets junior-level undergraduates at a major European research university. 
The curriculum has been refined over nine iterations (2017–2025), with several alumni subsequently pursuing careers in the \ac{HRE} field.
By reflecting on the evolution of the course organization, content, and assignments, we derive key lessons learned.
We further distill these insights into actionable design priorities for educators developing courses in rapidly evolving technological domains, emphasizing iterative growth and sustainable workload management for both students and instructors.
\end{abstract}




\maketitle

\section{Introduction}

\acfp{IC} are found in nearly every electronic device, from phones to medical equipment and systems supporting critical infrastructure.
Despite their ubiquity, the design and manufacturing of secure \ac{IC}s remains challenging as they are vulnerable to threats from malicious actors within the supply chain~\cite{guin2014counterfeit, halak2021cist, nygaard2024ethical}.
Hardware Trojans -- such as remotely triggered kill switches -- may be inserted at various points along the increasingly complex hardware supply chain, potentially compromising entire systems~\cite{DBLP:journals/dt/TehranipoorK10a, Bhunia2014_HwTrojan, Becker2019}. 
\acf{HRE} provides a means to detect such threats and ensure the security and trustworthiness of \acp{IC}~\cite{10318020, nygaard2024ethical}.

\ac{HRE} is the practice of analyzing an existing hardware to infer its design and specifications in the absence of original design information~\cite{rekoff1985reverse}. 
Acquiring this skill requires specialized training and education~\cite{cohen2018need}, yet opportunities to do so are scarce.
A recent systematic literature review of hardware security education identified only three dedicated \ac{HRE} courses worldwide~\cite{DBLP:conf/sigcse/WalendyW0PR25}, despite significant investment in semiconductor education from the EU and the US under initiatives such as the CHIPS Act~\cite{EU2023_ChipsAct, Ryan2022_CHIPSAct}.
This shortage is unsurprising given that \ac{HRE} is a highly specialized topic in a rapidly evolving technology domain. 

A central challenge in rapidly evolving technological domains is the lack of established blueprints or textbooks for course development.
Even when such resources exist, they quickly become outdated \cite{Lavicza2022DevelopingAE}, failing to capture the latest advances in academia and industry. 
To address this educational gap in \ac{HRE} and similar rapidly evolving fields, this paper, draws on eight years of experience teaching a  14-week, junior-level undergraduate \ac{HRE} course taught at a major public research university.

This experience report pursues two complementary objectives: 
(1)~to critically reflect on the evolution of our \ac{HRE} course over the past eight years, and 
(2)~to outline design priorities for educators developing courses in rapidly evolving technological domains.
To address the first objective, we reflect on longitudinal changes to the course organization, content, and assessment, informed by discussions with all current and former educators.
These reflections highlight that sustaining a hands-on course in a rapidly evolving technology domain requires continuous trade-offs between pedagogical ambition, institutional constraints, and educator workload.
To address the second objective, we distill lessons learned from this evolution into a small set of actionable design priorities and situate them within established educational guidance.

\section{Current Course Implementation}
Here, we describe the current structure and implementation of the \acf{HRE} course as realized in its most recent iteration, serving as the reference point after eight years of continuous development.

\subsection{Course Structure}
The course is designed as a 14-week program with an average enrollment of approximately 25 students. It is intended for junior-level undergraduate students majoring in cybersecurity, a discipline that integrates concepts from computer science and engineering with a focus on security.
As of the 2024/25 winter term, it carries 5~ECTS and consists of two weekly 90-minute instructional slots comprising 16 lectures and 10 interactive sessions, including project helpdesk and exam preparation.
Additionally, two guest lectures from industry and government partners complement the academic content with real-world perspectives.
In parallel with the lectures, students complete six take-home, hands-on projects, each supported by one or two dedicated helpdesk sessions for technical assistance.
The course concludes with a Q\&A session to prepare students for a 120-minute written exam.
The exam assesses students' \ac{HRE} knowledge, including their ability to reproduce procedures, reorganize knowledge, and transfer skills to novel contexts.
The final grade is percentage-based, comprising up to 45\% from hands-on projects and up to 60\% from the written exam, including a 5\% bonus.
A minimum of 50\% is required to pass.

\subsection{Learning Goals}
The course emphasizes hands-on skills in netlist reverse engineering and analytical reasoning, complemented by a conceptual understanding of core \ac{HRE} principles.
By the end of the course, students understand the motivation, relevance, and applications of \acf{HRE}, as well as the complete end-to-end workflow from netlist extraction to netlist interpretation.
They are practically able to extract netlists from \ac{FPGA} bitstreams and \ac{IC} images and to perform in-depth structural and functional analysis of these netlists using the HAL framework~\cite{8306831, 10.1145/3310273.3323419}, enabling them to reconstruct circuit behavior and assess hardware security properties.
While students do not operate physical equipment to produce \ac{IC} images, they understand the underlying processes and constraints.

\subsection{Course Content}
To support these learning goals, the course begins with the motivation and application domains of \acf{HRE}%
\footnote{For an overview of \ac{HRE} domains, see existing surveys and related literature~\cite{Torrance2011TheSI, mssassi2024fpga, azriel2021survey}}.
It then introduces fundamental hardware design concepts, including \ac{IC} and \ac{FPGA} design flows, hardware description languages, and essential open-source tooling.
Students are guided through \ac{IC} reverse engineering, covering sample preparation, imaging, image analysis, memory extraction, and netlist extraction from \ac{IC} images, with brief exposure to \ac{PCB} reverse engineering.
\ac{FPGA} reverse engineering is also covered, focusing on bitstream extraction, breaking bitstream encryption, reverse engineering bitstream formats, and converting bitstreams to netlists.
The central part of the course is netlist reverse engineering, addressing structural and functional analysis of control and data paths.
The content is complemented by coverage of state-of-the-art hardware security threats and defenses, including hardware Trojan detection and obfuscation techniques and corresponding attacks.
These topics are reinforced through hands-on projects that focus on gate-level netlist reverse engineering using HAL, while also covering netlist extraction from \ac{IC} images, \ac{FPGA} reverse engineering, and selected hardware design fundamentals such as logic synthesis.
For additional details on course content, we refer to the works of Wiesen, Becker, Walendy, \etal~\cite{DBLP:conf/sigcse/WalendyW0PR25, Becker2022, wiesen2018teaching, 9028668}.

\subsection{Course Educators}
The course was originally developed by a team of three researchers, including one with a PhD and two doctoral candidates.
Over its eight-year evolution, the teaching team has undergone several transitions, with one founding educator remaining involved since the first iteration.
In total, six educators have taught the course, with the current teaching staff comprising three, two holding PhDs and one in the final year of doctoral studies.
All educators are active researchers in \ac{HRE} and publish their work at leading security conferences.
Beyond the undergraduate course, current and former educators also teach condensed versions to industry and government participants, extending the course’s impact beyond the university.

\section{Method}
\label{sec:methods}
This experience report pursues two primary objectives:
(1)~to critically reflect on the course's evolution over the past eight years, and 
(2)~to derive design priorities for educators developing courses in rapidly evolving technological domains. 
Below, we briefly describe the approach we followed to achieve these objectives.

\subsection{Analysis \& Reflection on Course Evolution}
To systematically map changes to the course, two researchers conducted an extensive audit of internal course archives -- including slide decks and project materials -- covering all nine course iterations from winter term 2017/18 to winter term 2024/25.
The changes identified in this analysis were grouped into three categories:

\begin{itemize}[itemsep=0.1em, topsep=0.1em]
\item \textbf{Evolution of Course Setup:} Changes in course structure and examination formats.
\item \textbf{Evolution of Course Content:} Changes of the topics covered and their emphasis over time.
\item \textbf{Evolution of Course Assignments:} Changes in assignment formats and the skills they targeted.
\end{itemize}

To understand the reasons behind the observed changes, the three categories were condensed into timelines highlighting the most significant modifications between course iterations. 
All six educators who had taught the course then discussed these changes in a Zoom session and jointly recorded their reflections on a Miro board.
This process yielded 78 written responses, comprising 26 reasons for organizational changes, 30 for content changes, and 22 for assignment changes.
An independent researcher, who had not participated in teaching the course, subsequently summarized the responses and identified overarching themes within each category.
The results of these analyses are presented in \autoref{sec:results}.

\subsection{Derivation of Design Priorities}
To derive priorities for researchers designing courses in rapidly evolving technological domains, two authors jointly synthesized the themes emerging from the course evolution analysis and educator reflections.
Through iterative abstraction, they identified recurring patterns across course iterations and design decisions, condensing them into a small set of higher-level, actionable design priorities that are robust to technological change.

As a final step, we mapped the priorities to the \ac{4CID} model~\cite{vanmerrienboer2019fourcomponent} and the 12 Principles of Computing Pedagogy~\cite{raspberrypifoundation2021big} to situate them within existing educational theory and practitioner guidance.
The \ac{4CID} model by~\citet{vanmerrienboer2019fourcomponent} describes four essential components for teaching complex skills that transfer well into new situations such as the workplace.
Learners proceed through a series of realistic training tasks comprising problem solving, reasoning, and routine skills alike, while being supported with the applicable theoretical context as well as concrete procedural guidance.
While \ac{4CID} is domain-agnostic, the \ac{12P} established by the National Centre for Computing Education primarily target computing and programming education and provide concrete, practice-oriented guidance for educators.
Although coding is not the primary learning goal of our course, several of these principles translate well to broader course design.
The synthesis of both educational models and the priorities proposed are presented in \autoref{sec:recommendations}.

\section{Course Evolution}
\label{sec:results}
Based on changes in course materials and qualitative insights from educators, we present the evolution of the course across iterations \textbf{I1} (winter term 17/18) through \textbf{I9} (winter term 24/25), along with their underlying motivations.
The observed developments fall into three categories:
(i)~course setup, (ii)~course content, and (iii)~course assignments.
For each category, we distill key lessons learned.
To triangulate this educator-centered perspective, we additionally incorporate student perspectives on course effectiveness.

\subsection{Evolution of Course Setup}

\subsubsection{Workload and Credit Allocation}
\label{sec:organization:workload}
The course evolved from an ungraded, practice-oriented format awarding 3 ECTS in iterations \textbf{I1} and \textbf{I2} into its current 14-week, 5 ECTS graded format.
The first structural change followed the discovery that students lacked the theoretical foundation required for complex practical tasks.
This lack of knowledge caused the actual student workload to significantly exceed the assigned credits.
To address this gap while being restricted by regulations that capped credits for the initial course type, educators added a second curriculum to create two sequential courses for iteration \textbf{I3}.
The new course \textbf{C0} used a graded, lecture-based format awarding 5 ECTS for six weeks of theoretical instruction, followed by the practice-based course \textbf{C1} which awarded 3 ECTS.
However, this separate structure allowed students to earn most of the credits by attending only the theoretical course \textbf{C0}, which led to incomplete learning as they skipped the essential practical tasks in \textbf{C1}.
To ensure students also acquired practical skills, educators introduced a unified course in \textbf{I4} by reducing the combined 8 ECTS of the two previous modules to a single 5 ECTS allocation.
To compensate for this reduction in credits, the total workload was decreased in iteration \textbf{I5}.

\begin{lesson}
When students lack the necessary prerequisite knowledge, their workload for tasks increases.
If these gaps remain unaddressed, students must fill them independently. 
While addressing knowledge gaps through additional lectures does not eliminate the workload, but it does allow for workload to be redistributed into formal teaching time.
The additional lectures should be reflected in credit allocation.
\end{lesson}

\subsubsection{Exam Format}
In \textbf{I1} and \textbf{I2}, where no final exam was required by regulation, students needed to earn at least 75\% of the points on each project to pass.
For \textbf{I3} the course \textbf{C0} was graded and a final exam thus became mandatory. 
The exam format was written, and students could earn up to 10\% of bonus points through pen-and-paper exercises during the semester.
After \textbf{C0} and \textbf{C1} were merged in \textbf{I4}, the final exam remained mandatory, and project assignments became a major part of the final grade.
Students could earn up to 45\% of their final grade through project submissions and up to 60\% through the final exam, which included a 5\% bonus.

During the COVID-19 pandemic, fully remote teaching in \textbf{I5} required administering the final exam as a virtual oral assessment, which led to grade inflation.
Recognizing that oral and written exams did not adequately assess hands-on competencies, the educators decided in \textbf{I6} to instead use a longer final project accompanied by a short student presentation.
While this allowed for the assessment of hands-on skills, it significantly increased educator workload, raised administrative concerns about providing fair retake opportunities for students who failed, and made preventing cheating -- such as copying other students' solutions or receiving help from others -- impractical.
Ultimately, these limitations necessitated a return to a traditional written exam format in \textbf{I9}.

\begin{lesson}
Although project-based final exams can assess hands-on skills more effectively than traditional formats, they may introduce substantial and potentially unsustainable educator workload, as well as practical challenges related to cheating prevention and administering fair retake opportunities. 
A hybrid approach combining semester-long project work with a final written exam can strike a sustainable balance between valid skill assessment, academic integrity, and administrative feasibility.
\end{lesson}

\subsection{Evolution of Course Content}

\subsubsection{Closing Foundational Knowledge Gaps}
As mentioned in~\autoref{sec:organization:workload}, \textbf{I1} and \textbf{I2} revealed that many students lacked the necessary background knowledge in key areas.
These deficiencies hindered their ability to follow the course material and to engage with hands-on tasks effectively.
Students either needed reminders on topics they had previously encountered but did not fully retain -- such as interpreting transistor diagrams -- or lacked prerequisite knowledge, \eg, from hardware design.
The latter raised two concerns about the curriculum:
First, although our institution offered classes on hardware design fundamentals, they were not part of the cybersecurity curriculum, making them ineligible as enrollment prerequisites for our course.
Other classes that focus on hardware design in a security context were part of the curriculum but were scheduled to run in parallel with our course.
Impractically tight coordination between the courses would have been required to ensure that topics were covered in a meaningful order.
After identifying these underlying causes, prerequisite knowledge was introduced in \textbf{C0} of \textbf{I3} to make the course more self-contained.
Foundational theoretical concepts are first refreshed or newly introduced and then connected to the \ac{HRE} context.
Immediately following the lecture, students apply this knowledge in hands-on tasks.

\begin{lesson}
Even advanced students may lack the foundational knowledge needed for specialized hands-on tasks.
In cybersecurity education, educators may encounter this issue particularly in more niche areas like hardware, and less so in a software domain.
Sequencing theoretical instruction before hands-on activities is essential to ensure effective learning.
\end{lesson}

\subsubsection{Expanding the Scope of Netlist Extraction.}
Though the course focuses on netlist reverse engineering, it has consistently included how these netlists are extracted from \acp{FPGA}.
The newly introduced lecture in \textbf{I3} expanded this scope by describing physical netlist extraction from \acp{IC}. 
This expansion aimed to equip students with versatile skills and knowledge applicable across different hardware platforms.
As netlist reverse engineering techniques are mainly technology-agnostic, adding \ac{IC} reverse engineering naturally complemented the existing course material.

Unlike \ac{IC} concepts, \ac{FPGA} reverse engineering has been part of the course since its inception for two main reasons: 
(i) the availability of open-source \ac{FPGA} tools made the topic broadly accessible, and 
(ii) the educators already possessed relevant domain expertise and were actively conducting research in this field. 
These factors made integrating \ac{FPGA} reverse engineering relatively straightforward, as it did not require extensive additional resources.

In contrast, broadly incorporating \ac{IC} reverse engineering became feasible only after three developments converged: 
(i) the expansion of our network providing us with access to \ac{IC} RE knowledge and tooling, 
(ii) increased academic research activity in \ac{IC} reverse engineering, which supplied both methodological guidance and reference material,
and (iii) growing educator engagement with \ac{IC} technologies, which built the necessary in-house expertise to effectively teach the topic.

Supported by these advancements, \ac{IC}-related content could be gradually introduced between \textbf{I3} and \textbf{I8}, and lectures on obtaining and analyzing \ac{IC} images were added in collaboration with external \ac{HRE} laboratory partners.
New assignments were first introduced in the form of paper-based exercises on \ac{IC} standard-cell reverse engineering.
By \textbf{I5}, the assignments transitioned to digital cell reverse engineering using software tools licensed to us for educational use from research partners.
These tools also enabled students to extract netlists from the \ac{IC} images and analyze them within the HAL environment.
Thus, \ac{IC} netlist reversing engineering became an increasingly central part of the course.

\begin{lesson}
Hands-on skill acquisition in rapidly evolving technology domains benefits from access to appropriate tools. 
Starting with accessible tools, preferably open-source, that align with the educators' expertise provides a feasible entry point.
The toolset can then be gradually expanded through collaborations, partnerships, and engagement with current research.
\end{lesson}

\subsubsection{Integrating Current Research}
The course has continuously evolved to incorporate contemporary research developments, ensuring that students remain familiar with the latest tools, techniques, and security challenges in \ac{HRE}.  

Notable examples include the datapath reversing technique\linebreak \textit{DANA}~\cite{albartus2020dana}, introduced in \textbf{I5}, and the case study \textit{Maggie}~\cite{klix2024stealing}, introduced in \textbf{I8}.
In Maggie, an \ac{FPGA} embedded in an iPhone is reverse engineered, highlighting the real-world relevance of the course content.  
Due to the limited amount of material that can be covered within a 14-week schedule, new research topics such as bitstream manipulation on \acp{FPGA} necessitated the removal of older content, such as a watermarking scheme originally introduced in \textbf{I3}.

\begin{lesson}
Integrating current research and case studies into the curriculum is possible and drives student interest.
However, educators must resist the temptation to simply keep adding materials -- pruning older material is unavoidable to prevent scope creep and maintain a manageable workload for students.
\end{lesson}

\subsection{Evolution of Course Assignments}

\subsubsection{Student-Driven Transition to Python}
Students reported having difficulty with the C++ language used in HAL-based projects, and requested Python support in HAL.
Python rapidly became the language of choice for scripting in \textbf{I3}, and has remained a core component of take-home projects ever since.
Student feedback and bug reports have also guided the expansion and continuous improvement of HAL and its documentation in many other cases.
Besides refining HAL for the next course iteration, this feedback is a valuable resource for continuously improving HAL for the broader community.

\begin{lesson}
Open-source tools and student-driven feedback can create a powerful iterative improvement loop that enhances the usability and reliability of tools for students and the broader community.  
In addition, using a high-level language such as Python allows students to focus on domain-specific challenges rather than low-level programming syntax, facilitating more effective learning.
\end{lesson}

\subsubsection{Fostering Independent Learning and Hands-On Skills}
The course evolved in pursuit of the following two interconnected goals: (i) to foster students' ability to learn and work independently, and (ii) to develop hands-on skills relevant to real-world reverse engineering challenges.
To this end, two major changes were implemented in \textbf{I5}.
To achieve the first goal, educators gradually decreased instructional support throughout the semester to avoid providing excessive guidance.
To achieve the second goal, exercises that focused solely on theoretical concepts without demonstrating practical application were replaced with realistic tasks in the HAL framework.
This change was only possible due to significant improvements in HAL, including extended features and better usability.
Previously, these limitations restricted the complexity of hands-on tasks.
With increasing complexity of netlist reverse engineering tasks, students were also given more time.

\begin{lesson}
When tools are reliable, theoretical exercises can be replaced with hands-on tasks to foster independent learning and develop skills applicable to real-world challenges.
\end{lesson}

\subsection{Student Perspective}
Although this evaluation focuses on the educators' perspective on course evolution, it is essential to incorporate the student perspective when assessing its effectiveness.
Across all iterations, the course has maintained a high exam pass rate of approximately 93\%, and a retention rate of about 75\%, with roughly 260 students initially enrolled.
In voluntary institutional course evaluations, students have consistently rated the overall course and its teaching materials -- including lectures and exercises -- above four on a five-point scale (1 = very poor, 5 = very good) for their usefulness in learning the material, further supporting the course's effectiveness.
These quantitative metrics are also reflected in the long-term career trajectories of alumni.
Several former students have pursued doctoral research in hardware security, while others have transitioned into professional roles as \ac{HRE} experts.
Together, this evidence suggests that the course effectively bridges the gap between academic theory and sought-after skills in the semiconductor security sector.
\section{Priorities for Course Design on Rapidly Evolving Technology}
\label{sec:recommendations}

Designing -- and maintaining -- curricula in rapidly evolving technological domains within and beyond cybersecurity requires particular care.
In the following, we present three design priorities, informed by eight years of teaching \ac{HRE}.
These priorities are intended to support educators in addressing both curricular design and the administrative and organizational realities of sustaining such courses.
Where applicable, we contextualize these priorities using the \ac{4CID} model~\cite{vanmerrienboer2019fourcomponent} and the 12 Principles of Computing Pedagogy~\cite{raspberrypifoundation2021big}. Both are introduced in \autoref{sec:methods}, and the latter is hereafter abbreviated as \acused{12P}\ac{12P}.

\subsection{Start Small and Practical}
\label{sub:recommendations:startsmall}
In rapidly evolving domains such as \ac{HRE}, established blueprints for course design or even textbooks may not yet be readily available.
Moreover, educators cannot assume that comprehensive theoretical work exists, nor that all instructors have first-hand experience with all relevant aspects of the domain.
Under these conditions, a theory-first approach to course design might not only be suboptimal from a pedagogical perspective, it might not at all be practical.

% assignments
Instead, course contents should be grounded in what is currently reliable: concrete problems drawn from real-world practice.
We recommend carefully selecting a small number of such problems for which substantial expertise and experience exists within the team.
Deriving project-based assignments from these real-world problems will yield hands-on learning tasks that educators can confidently teach and realistically critique students' performance in.
Such assignments allow students to build advanced reasoning and problem-solving skills.
They also make abstract concepts tangible, aligning with \ac{12P}'s recommendation to \enquote{Make Concrete}.

% lecture
Naturally, students will need supporting knowledge to be able to work productively on these assignments.
We advocate conveying this knowledge through lectures that accompany the projects and scheduled immediately before the corresponding assignments.
This avoids splitting the curriculum in a purely theoretical and a purely practical part and instead tightly couples concepts and application.
This structure closely reflects the principles of the \ac{4CID} model: realistic learning tasks form the foundation of the curriculum, while theoretical content is introduced in direct support of those tasks rather than as an end in itself.
Selecting which concepts to teach requires a thorough analysis of the competencies required in each task.
At the same time, educators should be mindful of the students' potentially diverse levels of prior knowledge.
The curriculum should explicitly identify and close critical knowledge gaps, ensuring that all students can meaningfully engage with the assignments.
Finally, assessment methods should focus primarily on students' practical performance, not merely on measuring their retention of contextually detached theory.

\subsection{Iteratively Expand with Domain Experts}
\label{sub:recommendations:expand}

Having established an initial curriculum, subsequent iterations then call for updates for two reasons:
First, starting small yields opportunities for introducing additional content that was identified as relevant.
It is often impossible to know in advance all important aspects, particularly in fields which are highly multidisciplinary---such as in our case of \ac{HRE}.
Second, knowledge in the field evolves rapidly, making it impossible for a curriculum to remain final.

These challenges can be addressed by \textit{gradually} incorporating new material and collaborating with domain experts from academia and industry, who help identify relevant knowledge.
Such collaboration further ensures that hands-on tasks remain aligned with real-world challenges.
Our experience has shown that material based on current research is valuable and can be safely integrated into a course without overwhelming students with complexity, so long as overall workload is carefully managed.

\subsection{Manage Workload and Scope Creep}
\label{sub:recommendations:workload}

Every year new research findings, new attacks, and new countermeasures emerge.
Incorporating these important insights into the course is essential to maintain its currency and relevance.
However, this inevitably introduces a structural problem of scope creep. 
Both \ac{12P} and \ac{4CID} call on cognitive load theory and caution that students' working memory has inherent limits.
Thus, while including all the latest research in the course is tempting, doing so may prove counter-productive in practice.
Students may become cognitively overwhelmed, impairing their ability to absorb and process the amount of information. 
Furthermore, each additional concept consumes educator bandwidth, and there is a practical limit to how much content can be effectively prepared, taught, and graded.

Therefore, careful management of course scope is required to protect students and educators alike.
Removing or replacing content is necessary to create space for new material.
Conversely, compressing existing content to accommodate new material can remove essential theoretical knowledge needed to successfully perform complex hands-on tasks.
Additionally, increasing the number of topics addressed through hands-on assignments requires shortening projects, thereby reducing their depth conflicting with the \ac{4CID} model, which calls for coherent, \enquote{whole} problems.

\section{Discussion}

\subsection{Related Work}
A recent 2025 systematic literature review of hardware security education identified only three dedicated \ac{HRE} courses worldwide~\cite{DBLP:conf/sigcse/WalendyW0PR25}.
While information about course setups is available -- lectures combined with hands-on projects appear common practice -- there is limited public documentation of how these three courses have evolved and sustained over time. 
This experience report provides such a longitudinal perspective, documenting nine iterations of one of the three dedicated \ac{HRE} courses from the educator's viewpoint.

An earlier publication laid the pedagogical foundation for the course, proposing educational guidelines for practice-oriented \ac{HRE} courses, emphasizing the integration of graphical and textual representations of netlists alongside instructional support for developing conceptual and perceptual competencies~\cite{wiesen2018teaching}. 
\blind{Wiesen \etal}'s evaluation of \textbf{I1} identified significant knowledge gaps in foundational hardware concepts, prompting a recommendation to restructure the curriculum with dedicated theoretical instruction before hands-on tasks. 
Our longitudinal analysis of nine iterations reveals how those initial guidelines were operationalized and refined in practice, while the sustained 93\% pass rate and alumni career outcomes confirm the pedagogical approach's effectiveness at scale.

Building on \blind{Wiesen \etal}'s pedagogical foundation, a subsequent empirical study examined the cognitive factors underlying \ac{HRE} skill acquisition across 38 students from \textbf{I2} and \textbf{I3}~\cite{9028668}. 

The findings showed that working memory and processing speed are fixed cognitive constraints affecting performance. In contrast, students' prior knowledge in cryptography and finite state machines significantly predicts success -- and crucially, these knowledge gaps are remediable through structured instruction. 
This distinction between unchangeable cognitive limitations and addressable knowledge gaps directly shaped the course's evolution, with foundational lectures added before each assignment to establish the requisite knowledge background.

\subsection{Transferability to Other Teaching Contexts}
The course primarily relies on the open-source HAL environment and includes a self-contained theory component that provides all required background.
As no physical hardware (\eg, \acp{FPGA}) is required, it is highly portable across teaching contexts.
It has been delivered in several contexts beyond its original setting, including at a different institution in the US.
When time is limited, the lecture and projects can be restricted to either \ac{FPGA} or \ac{IC} reverse engineering.

In addition, the course has been adapted into two alternative formats, both delivered several times: (i)~a four-hour tutorial centered on a single project, offered at a summer school on real-world cryptography for early-stage PhD students; and (ii)~a three-day hands-on course targeting professionals seeking to extend their \ac{HRE} skills.
In both adaptations, Python programming skills proved to be a mandatory prerequisite, and prior hardware experience substantially improved learning outcomes.

While intellectual property restrictions imposed by laboratory partners prevent a fully open-source release, we make the \ac{HRE} materials available for academic use upon request.
They provide a concrete foundation for institutions seeking to bridge the gap between existing educational programs and industry practice, either by adopting the course in full or by adapting individual components to local needs.

\section{Conclusion}
In this eight-year longitudinal report, we reflect on the evolution of our \ac{HRE} course and outline priorities for creating courses in rapidly evolving technology fields.
The course design demonstrates how ECTS regulations and mandatory exams can be reconciled with an emphasis on hands-on projects that support workplace readiness.
Our experience shows how assignments can progressively evolve into complex, real-world tasks, while emerging theoretical gaps among students are addressed through targeted instruction.
The course evolution further illustrates how new research insights and topics, such as \ac{IC} reverse engineering, can be integrated while carefully managing the growing volume of teaching material.
We conclude that such curricula can never be final or exhaustive without overloading students and instructors, underscoring the need for deliberate strategies to prevent scope creep.

From these lessons, we distilled actionable design priorities for courses in rapidly evolving technology domains and situated them within the \ac{4CID} framework and the 12 Principles of Computing Pedagogy.
Specifically, educators should define tasks that closely resemble real-world workplace activities and introduce theoretical concepts in direct support of those tasks.
Curriculum evolution should be predominantly substitutive rather than additive, ensuring that new material replaces existing content to preserve assignment depth while preventing student overload.

Overall, this work offers practical guidance for educators seeking to design sustainable, hands-on courses in rapidly changing technology fields.


\begin{acks}
We thank all current and former educators -- Nils Albartus, Marc Fyrbiak, Simon Klix, and Julian Speith -- for their help in understanding the changes that have occurred in the course over the years.
We further thank Sarah Naqvi for her help with analyzing the course materials.
This work was supported by the Deutsche Forschungsgemeinschaft (DFG, German Research Foundation) under Germany’s Excellence Strategy -- EXC 2092 CASA -- 390781972, and by the \href{https://rc-trust.ai}{Research Center Trustworthy Data Science and Security}, one of the Research Alliance Centers within the \href{https://uaruhr.de}{UA Ruhr}.
\end{acks}

\bibliographystyle{ACM-Reference-Format}
%\balance
\makeatletter
\if@ACM@anonymous
    \bibliography{bibliography,own_bib_blinded}
\else
    \bibliography{bibliography,own_bib_clear}
\fi
\makeatother

%\appendix

\end{document}